# *Ab initio* calculations of the physical properties of transition metal carbides and nitrides and possible routes to high-$T_c$

E. G. Maksimov[(1)], S. V. Ebert[(1)], M. V. Magnitskaya[(2)],
A. E. Karakozov[(2)], S. Y. Savrasov[(3)]

[(1)] *P.N. Lebedev Physical Institute, Leninskii prosp. 53, Moscow, 119991 Russia*
[(2)] *Institute for High Pressure Physics of RAS, Troitsk, Moscow Region, 142190 Russia*
[(3)] *Department of Physics, University of California, Davis, California, 95616 USA*

*Ab initio* linear-response calculations are reported of the phonon spectra and the electron-phonon interaction for several transition metal carbides and nitrides in a NaCl-type structure. For NbC, the kinetic, optical, and superconducting properties are calculated in detail at various pressures and the normal-pressure results are found to well agree with the experiment. Factors accounting for the relatively low critical temperatures $T_c$ in transition metal compounds with light elements are considered and the possible ways of increasing $T_c$ are discussed.

## I. INTRODUCTION

In the 1970s and 1980s, the superconducting properties of transition metal carbides and nitrides were intensively studied both theoretically and experimentally (see, e.g., Refs. 1, 2). In particular, these compounds were considered as candidates to high-$T_c$ superconductors. For example, the qualitative estimates[3–5] of the electron–phonon coupling constant based on the rigid muffin-tin approximation imply that $T_c \approx 30$ K might be expected for the compound MoN, because its calculated density of states at the Fermi level, $N(0)$, is highest among monocarbides and mononitrides[3–5]. However, these calculations were performed for stoichiometric MoN with a NaCl-type cubic structure. Unfortunately, this crystal structure has never been realized for a stoichiometric composition; it is found[6] to be stable only in case of MoN$_{1-x}$ with $0.61 \leq x \leq 0.75$. As is well known, the superconducting transition temperature $T_c$ drops dramatically in N-deficient nitrides (C-deficient carbides). In line with this statement, superconductivity, apparently, was not observed in crystalline molybdenum nitrides. There were a few attempts to obtain epitaxial thin films of molybdenum nitride in the cubic structure (see Ref. 7 and references therein). However, the films with the cubic structure reported in the literature do not exhibit expected high $T_c$ values. The reason for their critical temperature being noticeably below the theoretical prediction is unclear.

There is another important factor, besides high values of $N(0)$, that favors reaching relatively high $T_c$ in these as well as some other systems, namely, sufficiently high phonon frequencies. According to the Bardeen – Cooper – Schrieffer (BCS) theory, the maximum critical temperature $T_c$ of superconducting transition might be observed in metals possessing minimum atomic masses $M$, since this theory predicts that $T_c \sim \omega_{ph} \sim 1/\sqrt{M}$. For this reason, much attention was devoted for many years to the possible superconductivity of metallic hydrogen (see, e.g., Ref. 8). The results of our recent *ab initio* calculations[9] showed that metallic hydrogen at ultrahigh pressures ($p \sim 20$ Mbar) might actually be a superconductor with $T_c \sim 600$ K. Unfortunately, not only any practical use, but even the synthesis of superconducting metallic hydrogen at such pressures is impossible. The possibility of reaching high $T_c$ values was previously also considered for metal hydrides[10]. Recently, Ashcroft[11] returned to this problem in the context of the revival of the general interest in searching for new compounds exhibiting superconductivity in the intermediate temperature range. In addition to metal hydrides, of considerable interest from the standpoint of reaching high $T_c$ values are transition metal borides, carbides, nitrides, and oxides, because they are also characterized by rather high frequencies of optical phonons associated predominantly with the vibrations of light atoms B, C, N, and O. One such compound with a relatively high critical temperature ($T_c \sim 40$ K), is magnesium diboride $(MgB_2)$[12], in which superconductivity is primarily due to the interaction between optical vibrations and the electrons localized at boron sites.

It is well known that at high pressure non-transition-metal superconductors exhibit a decrease in $T_c$, while many of transition metals and their compounds exhibit an increase in $T_c$. This observation stimulated in the 1970s and 1980s extensive studies of pressure influence on superconducting properties of transition metal systems, in particular, carbides and nitrides. Both $N(0)$ and the phonon frequencies may change when pressure is applied to a material, and in this paper we consider, in particular, how high pressure affects these quantities.

Note that detailed *ab initio* calculations of the electron and phonon spectra and, especially, the electron–phonon interaction (EPI) for carbides and nitrides are not numerous, because such calculations are tedious and time-consuming. An earlier example is offered by the phonon spectrum calculation[13] for niobium monocarbide NbC, performed using the full-potential linear muffin-tin orbital method (FP-LMTO) as implemented in Ref.

14. Recently, we have also briefly reported[15] the results of *ab initio* calculation of NbC made using the same FP-LMTO program package[14]. In Ref. 15 we calculated the electron and phonon spectra and the EPI spectral densities at normal pressure and upon isotropic compression by 15% and 30% in volume. Additionally, two recent papers should be mentioned, where the lattice dynamics and the electron–phonon interaction for NbC have been studied using first-principles pseudopotential methods[16, 17]. In Ref. 16 several other transition metal carbides and nitrides have been also calculated[16].

This work pursues two goals. First, to carry out *ab initio* calculations of the phonon spectra of some transition metal monocarbides and mononitrides in the NaCl structure, with a special emphasis on NbC, which is the most extensively studied among these compounds in respect of phonon-related properties. In particular, we calculate the electron–phonon interaction of NbC and such properties as electrical conductivity, optical spectra, and superconductivity, and compare the obtained results with available experimental data. Second, given that a good agreement of theoretical and experimental data is obtained, to investigate, without making use of any models, the possible routes for reaching higher $T_c$ values in transition metal compounds with light elements.

## II. CALCULATION OF THE ELECTRON AND PHONON SPECTRA AND THE EPI SPECTRAL DENSITY

### A. METHOD OF CALCULATION

Methods of *ab initio* calculations used in this work are described in detail in Refs. 13 and 14, as well as in our review articles[18, 19]. For this reason, we will not describe technical details and only briefly mention some principle points. Here, the main attention is paid to the discussion of obtained results and the physics of phenomena that occur in transition metal carbides and nitrides.

For calculating the kinetic and optical properties we used an approach which we name the hybrid method. Within this approach, the electron spectrum of a metal is calculated using the density functional theory developed in Ref. 20 and 21, that is the Kohn – Sham equation is solved. The phonon spectrum, as well as the change in the effective potential due to the displacements of atoms are calculated using the linear response theory for the Kohn – Sham equation. With the electron and phonon spectra obtained, one is able to calculate the EPI matrix elements and spectral densities (Eliashberg functions). The next step consists in applying a many-particle perturbation theory based on Fröhlich Hamiltonian to determine the electronic properties of metal. The whole procedure is described in detail in Refs. 14, 18, 19, and 22, where the adequacy of this hybrid approach in calculating the EPI and its effect on the physical properties of metals is also discussed.

Our basic self-consistent calculations of the compounds NbC, YC, HfC, MoC, MoN, and NbN in the cubic structure were performed at a (32×32×32) reciprocal-lattice grid, which corresponds to 897 **k**-points per the irreducible Brillouin zone (IBZ). The von Barth–Hedin-like exchange-correlation formulas after Ref. 23 were employed. A multiple-$\kappa$ basis set was used. The calculations involved the following parameters: $3\kappa$–$spd$ LMTO basis per transition metal atom (27 orbitals) and $3\kappa$–$sp$ LMTO basis per light atom (12 orbitals). The one-center expansions inside the MT-spheres were performed up to $l_{max} = 6$. The charge densities, the potentials, and their changes under perturbation were represented by spherical harmonics up to $l_{max} = 6$ inside the MT-spheres and by plane waves with about 140 Ry energy cutoff in the interstitial region. The **k**-space integration was made by means of the improved tetrahedron method[24]. The dynamical matrix was calculated at the 29 irreducible **q**-points of the (8×8×8) grid. The **k**-space integration for the matrix elements was performed over a (8×8×8) grid, the same grid as for the phonon wave vectors **q**. In addition, the integration weights for the **k**-points of this grid have been found to take into account the effects arising from the Fermi surface and the energy bands precisely. This was done with the help of a denser mesh of 897 **k**-points per the IBZ, which corresponds to the same (32×32×32) grid as for the energy bands calculation. Our calculated phonon spectra for HfC, MoC, MoN, and NbN are generally in good agreement with those obtained in Ref. 16. The results for these compounds will be discussed in Section IV.

In the calculation of the dispersion curves of NbC along the high-symmetry directions, additional **q**-points lying on a (16×16×16) grid were calculated to better resolve the phonon anomalies. Now we turn to the description of obtained results for NbC.

B. THE RESULTS FOR NbC

The calculated equilibrium lattice parameter $a = 8.396$ a.u. is only by 1% less than the experimental value at room temperature. We calculated the electron and phonon spectra at the theoretical equilibrium volume. The total electron density of states (DOS), $N(E)$, for NbC is shown in Fig. 1a, and the partial DOSs at the Nb and C sites are shown in Figs. 1b and 1c, respectively. Also shown in Fig. 1a is the total DOS for compressions of $V/V_0 = $ 0.85 and 0.7 which correspond to pressures of about 60 and 150 GPa, respectively. At normal pressure, the DOS peak at an energy of ~0.2 Ry above the Fermi level, $E_F$, is due to the unoccupied 4$d$ states at the Nb site. The peak at ~0.3 Ry below $E_F$ is almost equally contributed by the carbon 2$p$ states and the filled Nb 4$d$ states. The shape of partial $s$, $p$, and $d$ DOSs at the Nb and C sites implies that the chemical bonding in NbC is substantially ionic: the Nb electrons are transferred to the carbon 2$p$ states lying 0.3 – 0.35 Ry below the Fermi level. Near $E_F$, the carbon $p$ states are only slightly hybridized with the Nb $d$ states, and only tails of the C $p$ states emerge at the Fermi level.

Here, we do not present a plot of calculated phonon spectrum for NbC, because both the phonon dispersions and phonon density of states coincide very well with those obtained earlier by Savrasov (see Fig. 2 in Ref. 13). A good agreement between calculated and measured[25] frequencies is found for the most part of the phonon spectrum. In line with Savrasov's results[13], we also obtained a negligible overlap of the acoustic and optical vibration modes. Calculation of the phonon polarization vectors demonstrates that the acoustic part of spectrum is formed mostly by the vibrations of heavier Nb atoms, while the optical part by the vibrations of light carbon atoms. The calculated phonon density of states consists of virtually isolated parts related to the spectra of acoustic ($0 \leq \omega \leq 8$ THz) and optical ($15 \leq \omega \leq 20$ THz) vibrations.

Generally, the phonon dispersions calculated using the first-principles pseudopotential method[16, 17] agree rather well with our results obtained by all-electron LMTO method. However, the authors of Ref. 16 notice that their calculated phonon spectrum of NbC reveals anomalies along both the [ξ 0 0] and [ξ ξ 0] high-symmetry directions, while in Savrasov's calculation[13] the anomaly along the [ξ ξ 0] direction is not so pronounced. The same is true for our calculated phonon dispersions at the mesh of 29 **q**-points, which coincides with Savrasov's results. We can add that in line with the results of Ref. 16, the phonon spectrum of NbC obtained in Ref. 17 also demonstrates such a more pronounced dip along the [ξ ξ 0] direction. It should be noted, however, that in our calculation all three acoustic modes at **q** ≈ [0.65 0 0] and two upper acoustic modes at [0.5 0.5 0] are more close to the experimental frequencies than in the pseudopotential calculations[16, 17], which is clear from comparison of Fig. 2 in Ref.13 with Fig. 2 in Ref. 16 and Fig. 2 in Ref. 17 (the experimental points are also displayed in the figures). To investigate the discrepancy in the low-frequency transverse acoustic mode at **q** ≈ [0.5 0.5 0], we carried out additional calculations using finer grids of **k**-points and **q**-vectors, and did find a dip in this mode at **q** ≈ [0.5 0.5 0] missed in 29-points calculation, although the dip is still not so pronounced as in Ref. 16, 17. We believe that this discrepancy between our results and the results of pseudopotential calculations[16, 17] can be due to different approaches used. In our case the phonon frequencies are calculated directly at all the **q**-points involved, while the authors of Refs. 16, 17 applied the force constant method, which might artificially enhance the peculiarity in question. We cannot state this with assurance, because in the literature there is no experimental information on this particular mode of NbC.

We calculated the Eliashberg function, or the EPI spectral density $\alpha^2(\Omega)F(\Omega)$, which describes the EPI effect on the single-particle properties of the electron system, including superconducting properties, as well as the transport spectral density $\alpha^2_{tr}(\Omega)F(\Omega)$, which describes the kinetic properties in the system of electrons interacting with phonons. The calculated functions $\alpha^2(\Omega)F(\Omega)$ at normal pressure and at 15% and 30% compressions are shown in Fig. 2 by solid, dashed, and dash-dotted lines, respectively. We do not display the function $\alpha^2_{tr}(\Omega)F(\Omega)$ because it looks very similar to the Eliashberg function $\alpha^2(\Omega)F(\Omega)$. This is also the case with most of the metals we have studied before[18, 19]. These two functions should be substantially different only for systems with strong anisotropy of EPI. As is seen in Fig. 2, the EPI spectral density consists of two parts significantly spaced on the frequency scale, in accordance with the shape of our calculated phonon density of states (not presented). The low-frequency part describes the interaction of electrons with acoustic phonons, while the high-frequency part refers to the interaction with optical phonons, just as in the case of the phonon DOS.

### III. CALCULATION OF THE KINETIC, OPTICAL, AND SUPERCONDUCTING PROPERTIES OF NbC

The knowledge of the EPI spectral characteristics together with the Kohn – Sham spectra of electron excitations $\varepsilon_{k\lambda}$ allows one to calculate all the kinetic and superconducting properties, which, in our opinion, for NbC are mainly due to the EPI. The intraband frequency-dependent conductivity $\sigma_b(\omega)$ can be written in the following form:

$$\sigma_b(\omega) = \frac{\omega_{pl}^2}{4\pi}\left[-i\omega\frac{m_{tr}(\omega,\tau)}{m_b} + \frac{1}{\tau_{tr}(\omega,\tau)}\right]^{-1}, \quad (1)$$

where, $\omega_{pl}$ is the plasma frequency of the electrons:

$$\omega_{pl}^2 = \frac{8\pi e^2}{3v} \sum_{k\lambda} \left|\frac{\partial \varepsilon_{k\lambda}}{\partial \bm{k}}\right|^2 \delta(\varepsilon_{k\lambda} - \varepsilon_F) \quad . \tag{2}$$

Expression for the transport mass renormalization $m_{tr}(\omega, \tau)/m_b$ and transport relaxation rate, or the inverse lifetime $1/\tau_{tr}(\omega, \tau)$, has the form[22]

$$\omega \frac{m_{tr}(\omega,\tau)}{m_b} + \frac{i}{\tau_{tr}(\omega,\tau)} = \omega + 2\int_0^\infty d\Omega\, \alpha_{tr}^2(\Omega) F(\Omega) K\left(\frac{\omega}{2\pi\tau}, \frac{\Omega}{2\pi\tau}\right) + \frac{i}{\tau_{imp}} \quad . \tag{3}$$

Here,

$$K(x,y) = \frac{i}{y}\left\{\frac{y-x}{x}[\Psi(1-ix+iy) - \Psi(1+iy)] - y\right\} \quad , \tag{4}$$

where $\Psi(1+iy)$ is the digamma-function. Definition (3) of the relaxation rate includes the electron scattering by impurities described by the last term $\sim 1/\tau_{imp}$, which is a single adjustable parameter here, while all the rest in formula (3) is found from first principles. The value of $\gamma_{imp} = 1/\tau_{imp}$ was estimated from comparison of residual resistivity at $T = T_c$ with the experimental data. Our calculated electron plasma frequency $\omega_{pl}$ of NbC is equal to 7.57 eV. Figure 3 displays the temperature dependence of the electrical resistivity $R(T) = 1/\sigma(\omega = 0)$ obtained in our calculations and in the experiment[26]. Comparison of the theoretical and experimental absolute values of $R(T = T_c)$ gives for $1/\tau_{imp}$ a value of 0.18 eV. At high temperatures $T \geq 0.3<\omega>$ (where $<\omega>$ is a characteristic phonon frequency), the electrical resistivity due to the electron scattering by phonons can be written as

$$R = \frac{4\pi}{\omega_{pl}^2} 2\pi \lambda_{tr} T \quad , \tag{5}$$

where $\lambda_{tr}$ is the electron–phonon coupling constant

$$\lambda_{tr} = 2\int_0^\infty \alpha_{tr}^2(\Omega) F(\Omega) \frac{d\Omega}{\Omega} \quad . \tag{6}$$

Our estimate gives the value 0.92 for $\lambda_{tr}$. As is seen in Fig. 3, the slopes of theoretical and experimental dependencies are rather close to each other, with a deviation not exceeding a few percent. It should be taken into account that the experimental resistivity was measured for the non-stoichiometric compound $NbC_{0.98}$ that is close to but still different from the stoichiometric compound. Nevertheless, even in this case the residual resistivity is rather high, $\rho_0 = 24$ μΩ cm. For such a high value of $\rho_0$, Matthiessen's rule (used in formula (3)) of the simple additivity of the impurity and phonon relaxation times may itself be violated, which could result in some discrepancy between the experimental and theoretical data.

Indeed, such a discrepancy is seen in temperature range 100 K < $T$ < 200 K, where the experimental dependence is noticeably more convex in comparison with the theoretical curve. The authors of Ref. 26 attributed this deviation of temperature dependence of $R(T)$ from the linear behavior (5) due to the EPI to the occurrence of electron states on the vacant C sites in the system $NbC_{1-x}$. According to their suggestion, the energy of these states is very close to the Fermi level and thus they can affect the electrical resistivity, resulting in a deviation of $R(T)$ from linearity. Notice, however, that this deviation is very small.

We also calculated the optical spectra of NbC over a wide frequency range, up to $\omega \approx 34$ eV. To evaluate the contribution from interband electron transitions to the optical conductivity, we used the simple approximation of non-interacting band electrons:

$$\sigma_1^{\text{inter}}(\omega) = \frac{e^2}{3\pi m^2 \omega} \sum_{\lambda \neq \lambda'} \int_{BZ} d\mathbf{k} \left|\langle \mathbf{k}\lambda' | \nabla | \mathbf{k}\lambda \rangle\right|^2 \Theta(\varepsilon_F - \varepsilon_{\mathbf{k}\lambda}) \Theta(\varepsilon_{\mathbf{k}\lambda'} - \varepsilon_F) \delta(\varepsilon_{\mathbf{k}\lambda'} - \varepsilon_{\mathbf{k}\lambda} - \omega) , \qquad (7)$$

where $\sigma_1^{\text{inter}}(\omega)$ is the real part of the interband conductivity, $e$ and $m$ are the charge and mass of an electron. To numerically evaluate $\sigma_1^{\text{inter}}(\omega)$, we employed the full-potential linear augmented plane wave method (FP-LAPW) as implemented in the program package WIEN2k[27]. The imaginary part of the interband conductivity was restored using the Kramers – Kronig relations. Then we calculated the reflectance spectrum $R(\omega)$ using the obtained optical conductivity. Figure 4 shows the calculated and measured[26] reflectance. Considering that we did not use any adjustable parameters in this work, except for the impurity relaxation rate $\gamma_{\text{imp}}$, the agreement of the theoretical and experimental data is rather good. There is a discrepancy between the calculated and measured spectra in the energy range 1.5 eV < $E$ < 3 eV. It is rather difficult to reveal the true reason for this discrepancy, because there is no well developed technique for calculating the properties of non-stoichiometric crystals. Note that this discrepancy occurs at relatively high energies, therefore it does not effect the evaluation of the superconducting properties of NbC.

Figure 5 displays the real and imaginary parts of the optical conductivity, $\sigma(\omega) = \sigma_1(\omega) + i\sigma_2(\omega)$. Also shown is the experimental optical conductivity[26]. It is seen that our calculations describe the experimental spectrum rather well, namely, the theoretical positions of $\sigma(\omega)$ features are agree with the experiment. There is, however, some disagreement between the amplitudes of calculated and measured spectra $\sigma_1(\omega)$ and $\sigma_2(\omega)$, especially for the imaginary part $\sigma_2(\omega)$. One should remember that in the experiments, the reflectance $R(\omega)$ was measured, rather than $\sigma(\omega)$ proper. Experimentally, the optical conductivity was determined by means of the Kramers – Kronig procedure for the reflectance that was measured only to the energies $E = 11$ eV. Our results show that there exist intense interband transitions at energies higher than 11 eV, which were not taken into account in Ref. 26. These transitions are clearly seen in Fig. 5, where the theoretical $\sigma(\omega)$ is displayed up to 17 eV. This fact is of great importance for the amplitude of the imaginary part of $\sigma(\omega)$ and can lead to the discrepancy between experimental and theoretical results.

We used the calculated from first principles Eliashberg function $\alpha^2(\Omega)F(\Omega)$ to solve the Eliashberg equation[28] and evaluate the order parameter $\Delta(\omega)$, as well as the renormalization function $Z(\omega)$. We also calculated the electron DOS in the superconducting state, $N(\omega)$, and the tunneling characteristics. The Eliashberg equation includes, besides the EPI, the Coulomb pseudopotential $\mu^*$. Our estimate of the superconducting transition temperature $T_c$ made with neglect of the Coulomb repulsion $\mu^*$ (i.e., including only the EPI) gives the value 15.7 K. As is well known, the presently available methods of *ab initio* calculations do not allow the Coulomb repulsion $\mu^*$ to be estimated. For this reason, we have determined $\mu^*$ from the condition of coincidence of the calculated and experimental values of the critical temperature ($T_c^{exp} = 11.1$ K), which yields $\mu^* = 0.15$. This value is likely to be somewhat overestimated, since $\mu^*$ lies in the range 0.12 to 0.13 for the majority of standard metals. It should be noted, however, that in the experiment the value of $T_c$ was determined for to a some degree non-stoichiometric carbide $NbC_{1-x}$, rather than for stoichiometric NbC. As is mentioned above, $T_c$ of the transition metal carbides drops quite rapidly with increasing $x$. This can also be the reason for $\mu^*$ to be overestimated. Our estimate gives for the ratio $2\Delta/T_c$ the value 4.1, very close to the experimental value (~4), which implies the existence of strong coupling effects in NbC. Here, we do not describe in detail the calculation of the superconducting characteristics of NbC, in particular, the tunneling $I$–$V$ curve, because we do not know any related experimental data.

Now we briefly discuss pressure-induced changes in the superconducting properties of NbC. As is seen from Fig. 1, The electron DOS at the Fermi level slightly decreases upon compression. The acoustic and optical phonon frequencies increase with increasing pressure, an increase being considerably larger for the optical phonon modes (see Fig. 2). The electron–phonon coupling constant $\lambda$ decreases with compression, which leads to a decrease in $T_c$. Qualitatively, this result is in agreement with some preliminary experimental data for NbC at high pressure (see Ref. 29). Unfortunately, there is no detailed experimental information on how the phonon spectra and the critical temperature of NbC change under pressure.

## IV. DISCUSSION OF POSSIBLE ROUTES TO HIGH-$T_c$

The obtained results demonstrate that our hybrid approach that has been successfully employed for calculating the properties of elemental metals[19] is quite appropriate in the case of transition metal carbides, too. Note that we use the Fröhlich Hamiltonian to take into account the EPI only in the electron spectrum and do not renormalize the phonon frequencies, since our calculated adiabatic phonons correctly describe the experimentally measured vibration spectrum in the main part of the Brillouin zone.

Returning to the question of possible existence of relatively high $T_c$ in transition metal carbides, we can say the following. Standard BSC theory gives for $T_c$ the expression

$$T_c = 1.14\varpi \exp\left(-\frac{1}{\lambda - \mu^*}\right), \qquad (8)$$

where $\varpi$ is an average phonon frequency. The EPI constant $\lambda$ can be expressed as

$$\lambda = N(0)\langle V^2 \rangle, \qquad (9)$$

where $\langle V^2 \rangle$ is an average EPI matrix element.

According to formula (9), the coupling constant $\lambda$ is determined by the DOS at the Fermi level, $N(0)$. As is mentioned in Introduction, this is the reason why the authors of early papers[3–5] have considered the enhancement of $N(0)$ as a possible way for increasing $T_c$ in carbides. In carbides and nitrides with cubic structure, the electron DOS is well described in the rigid-band approximation[30], that is, by simply shifting the Fermi level position. In the compound ZrC, the number of electrons per unit cell is smaller by one than that in NbC and, accordingly, the Fermi level is situated approximately 0.15 Ry lower and falls within a DOS minimum (see Fig. 1). In agreement with this picture, ZrC is not a superconductor. For NbN whose number of valence electrons is greater by one than that of NbC, the Fermi level shifts rightward to higher DOS values. Accordingly, this compound, if existed in a stoichiometric composition, would be characterized by a higher transition temperature ($T_c$ = 17 K). In MoC and MoN, $E_F$ shifts rightward to an even greater extent, since the number of valence electrons per unit cell is greater respectively by one and two than that in NbC. The DOS at the Fermi level in MoN is almost twice as high as in NbC. In accordance with the rigid-band estimate[3], the transition temperature in MoN must be on the order of 30 K, although, as mentioned above, such 'high-$T_c$' superconductivity in MoN has not been observed experimentally. The point is that this compound in crystalline form was never obtained with the stoichiometric composition and real samples always exhibited a rather large deficiency of nitrogen. It was demonstrated[31] that the shear elastic constant $C_{44}$ in cubic MoN is negative, which is an evidence of the crystal instability. This fact was later confirmed in Ref. 30, where it was also pointed out that only the presence of nitrogen vacancies can stabilize the cubic B1-type structure of MoN. In the same paper[30], positive elastic constants were obtained for MoC, but an instability of the acoustic mode in the X point at the Brillouin zone boundary was found. The instability in MoC was recently proved[16] by the phonon spectrum calculation in the whole Brillouin zone. An even stronger instability around the X point was reported[16] to exist in MoN and NbN. We carried out direct linear-response calculations of the phonon spectrum for MoC, MoN, and NbN. Our results confirm the conclusion of Ref. 16 that in these compounds a transverse mode in the [ξ ξ 0] direction is unstable over a range of wave vectors **q** near the X point. This implies that higher $T_c$ can hardly be reached in cubic carbides and nitrides by means of increasing $N(0)$, since such an increase is most likely to result in the lattice instability of a stoichiometric compound.

As was already mentioned, another possibility for increasing $T_c$ is to involve the optical phonons related to the vibrations of light atoms (C, N, etc.) in the formation of a superconducting state. Our calculations described in Section IIB demonstrate that for NbC the function $\alpha^2(\Omega)F(\Omega)$ consists of two separate parts: the low-frequency peak related to the acoustic phonons and the high-frequency peak related to the optical phonons. Accordingly, one may define two EPI constants:

$$\lambda_{ac} = 2\int_0^{\omega_1} \frac{d\Omega}{\Omega} \alpha^2(\Omega)F(\Omega) \ ,$$

(10)

$$\lambda_{op} = 2\int_{\omega_1}^{\infty} \frac{d\Omega}{\Omega} \alpha^2(\Omega)F(\Omega) \ ,$$

where $\omega_1 = 14$ THz (see Fig. 2). The critical temperature of a superconductor with two isolated phonon peaks can be presented in the form[1]

$$T_c = \frac{\omega_{\log}}{1.4} \exp\left(-\frac{1+\lambda}{\lambda - \mu^*}\right) . \qquad (11)$$

Here, $\lambda = \lambda_{ac} + \lambda_{op}$; $\omega_{\log}$ can be written as

$$\omega_{\log} = \omega_{ac}^{\nu} \omega_{op}^{1-\nu} , \qquad (12)$$

where $\nu = \lambda_{ac}/(\lambda_{ac} + \lambda_{op})$.

Our estimate of these constants gives $\lambda_{ac} = 0.71$ and $\lambda_{op} = 0.21$. Thus, for NbC the constant of coupling with optical phonons, $\lambda_{op}$, is more than three times lower than $\lambda_{ac}$, and, in accordance with expressions (11) and (12), its contribution to $T_c$ is also small. The reason why $\lambda_{op}$ is small can be easily explained within rigid muffin-tin approximation used in Refs. 3 – 5. Within this approximation, the coupling constant $\lambda$ for binary compounds of *AB*-type with substantially different ion masses can be expressed as

$$\lambda = \lambda_{ac} + \lambda_{op} = \frac{N_A(0)\langle I_A^2 \rangle}{M_A \langle \omega_A^2 \rangle} + \frac{N_B(0)\langle I_B^2 \rangle}{M_B \langle \omega_B^2 \rangle} . \qquad (13)$$

Here, $\langle I^2 \rangle$ is the matrix element of the squared gradient of muffin-tin potential integrated over the Fermi surface and the product $\eta = N(0)\langle I^2 \rangle$ is the Hopfield parameter[32]. According to estimates of Ref. 4, carbides and nitrides obey the relation $M_A \langle \omega_A^2 \rangle \approx M_B \langle \omega_B^2 \rangle$. Qualitatively, smaller values of $\lambda_{op}$ are explained by both a low $N_C(0)$ and a small $\langle I^2 \rangle$. Thus, a more promising way to reach higher $T_c$ is the search of a carbide possessing such a position of chemical potential that $N_C(0)$ would be sufficiently high.

An example could be offered by yttrium monocarbide YC whose number of electrons per unit cell is lower by two than that in NbC. In the rigid-band approximation, the Fermi level in YC falls within a peak in the DOS at $E = -0.3$ Ry, a feature formed predominantly by the carbon *p*-states (see Fig. 1). Unfortunately, the compound YC with stoichiometric composition is still not synthesized. Nevertheless, we performed detailed *ab initio* calculations of hypothetical YC with the NaCl-type structure and obtained its electron and phonon spectra, as well as Eliashberg function. We found no instabilities in the phonon spectrum of cubic YC and the cause that hinders the synthesis of the stoichiometric yttrium carbide is still unclear to us. It is conceivable that some other phase of the yttrium–carbon system is energetically preferable at equilibrium conditions.

We solved the Eliashberg equation and estimated the possible $T_c$ values for YC. Contrary to the expectations, our calculated $T_c$ value turned out rather low, 9.2 K, which is even lower than for NbC. Let us consider possible reasons for this fact. Our results on the electron DOS for YC (Fig. 6) confirm the above observation that the rigid-band approximation is quite suitable for describing the electron states in monocarbides: both the total and partial DOSs of YC become very similar to those of NbC, as the chemical potential is shifted by –0.25 Ry on the energy scale. Our calculated phonon spectra and Eliashberg function for YC are shown in Fig. 7a and 7b, respectively. The acoustic phonons have no peculiarities and their frequencies are close in value to the acoustic frequencies in NbC. The optical phonons, however, are much softer and their average frequency is by factor of 1.5 smaller than for NbC (see Fig. 2). The acoustic and optical coupling constants for YC are also significantly different from the case of NbC. For YC, the acoustic constant $\lambda_{ac} = 0.19$ is much less than for NbC, while the optical constant $\lambda_{op} = 0.52$ is 2.5 times as large as $\lambda_{op}$ for NbC. Thus, the $T_c$ value in YC is mainly determined by optical phonons, which implies that it makes sense to search for superconductors whose optical phonon modes mainly contribute in $T_c$. Nevertheless, the absolute value of $\lambda_{op}$ for YC is insufficient to result in a high $T_c$.

Let us consider this question in more detail. In yttrium carbide, the DOS at the Fermi level, $N(0)$, is sufficiently high. Moreover, $N_C(0)$ in YC is even larger than $N_{Nb}(0)$ in NbC, but the optical coupling constant $\lambda_{op}$ is smaller than the acoustic constant $\lambda_{ac}$ in NbC. According to formula (13), the value of the EPI coupling constant is determined by the Hopfield parameter. The first moment of Eliashberg function can be expressed[33] as

$$2\int_0^{\omega_1} \omega\alpha^2(\omega)F(\omega)d\omega = \sum_i \frac{\eta_i}{M_i} \quad . \tag{14}$$

Here, $\eta_i = N_i(0)\langle I_i^2\rangle$ is the Hopfield parameter of an $i$th ion and $M_i$ is the ion mass. We calculated 1$^{st}$ moments for the acoustic and optical modes separately, as we did for the coupling constants $\lambda$. This made it possible to find, in accordance with formulas (13) and (14), the Hopfield parameters of transition element and carbon for NbC and YC. In the case of NbC

$$\eta_{Nb} = 8.3 \text{ eV Å}^{-2} \quad \text{and} \quad \eta_C = 3.5 \text{ eV Å}^{-2},$$

while for YC

$$\eta_Y = 1.7 \text{ eV Å}^{-2} \quad \text{and} \quad \eta_C = 2.9 \text{ eV Å}^{-2}.$$

Thus, in YC the Hopfield parameter of C atom, $\eta_C$, exceeds $\eta_Y$ and $N_C(0)$ also exceeds $N_Y(0)$. This results in $\lambda_{op}$ being higher than $\lambda_{ac}$, that is the electrons interacting with the optical vibrations of C atoms play a main part in the formation of superconducting state. Here, it is important that the absolute value of $\eta_C$ for YC is even less than for NbC, which is due to the fact that in YC the electron states near $E_F$ are almost pure carbon $p$ states with a small admixture of carbon $s$ states. The optical phonons in metal carbides can be considered, to a good accuracy, as local vibrations of C atoms. In this case the Hopfield parameter $\eta$ can be written to a good approximation as

$$\eta = N(0)\langle I^2 \rangle \approx \left| \langle \psi_i | \nabla V_{eff} | \psi_j \rangle \right|^2 (N_i(0) N_j(0)). \quad (15)$$

Here, the indices $i$ and $j$ denote the values of the orbital angular momentum $l$ and the matrix elements of the potential gradient are nonzero only when the orbital moments of the wave functions $\psi_i$ and $\psi_j$ differ by 1. The Hopfield parameter of YC is small because $N_s(0)$ and $N_d(0)$ are very low.

Since $\eta_C$ for YC is even less than for NbC, a rather high value of $\lambda_{op}$ in YC is due to a strong softening of the transverse optical mode in YC as compared to NbC. The effect of lattice softening on superconductivity has long been a subject of wide discussion (see, e.g., Ref. 1) and the optical phonon mode in YC is a prominent example how the phonon softening can assist in increasing $T_c$. We would like to emphasize that the higher value of $\lambda$ in YC as compared to NbC is due to the softening of entire optical part of the phonon spectrum. A somewhat similar situation, namely a strong softening of the entire bond-stretching mode, is believed[34, 35] to exist in MgB$_2$ and boron-doped diamond. It should be noticed that peculiarities of the phonon spectrum by no means always lead to an increase in $T_c$. For instance, it has been frequently stated, in the past[1] as well as now[16, 17], that the existence of superconductivity in group-V transition metal monocarbides, in contrast to group-IV metal monocarbides, is related to peculiarities in their phonon spectrum. It is easy to make sure by direct linear-response calculation of the phonon spectra that $M_M \langle \omega_{ac}^2 \rangle \approx$ const for both group-V and group-IV transition metal monocarbides. This means, in accordance with expression (13), that $\lambda_{ac}$, as well as $T_c$, is defined mainly by the Hopfield parameter, rather than by the phonon spectrum and its peculiarities. We also calculated the superconducting properties of compressed YC and found that, as in case of NbC, its EPI constant $\lambda$ and $T_c$ decrease under pressure.

Although our predicted $T_c$ for hypothetical YC is rather small, we would like to mention that these are yttrium carbides which exhibit highest $T_c$ values among compounds of this class. In particular, the superconducting transition temperature of complex carbide $Y_2C_3$ is 18 K[36], while in some yttrium borocarbides it reaches a level of 15 – 23 K[37]. Possible reasons for a large (more than 4 times) difference in critical temperature between the compounds $Y_2C_3$ and $YC_2$ ($T_c$ = 4 K) have been recently considered[38] on the basis of *ab initio* calculations. According to Ref. 38, the total DOS at the Fermi level, $N(0)$, in $Y_2C_3$ is by 40% higher than in $YC_2$. Moreover, $N_C(0)$ in $Y_2C_3$ is twice as large as in $YC_2$ and this fact can be responsible for the higher $T_c$ value in the former. On the other hand, the calculations of Ref. 38 demonstrate that in $Y_2C_3$ the partial DOSs $N_s(0)$ and $N_d(0)$ on the C site are much smaller than $N_p(0)$. This implies, in line with our above discussion of the role of Hopfield parameter in YC, that the case of $Y_2C_3$ may be not optimal from the viewpoint of reaching high $T_c$ values. Estimations of Ref. 39 show that a strongest EPI in $Y_2C_3$ is related to the acoustic modes. The interaction with the bond-stretching mode is rather weak because of a high frequency of this mode. It is assumed[39] that this interaction can be enhanced by going to a higher electron count. However, it is unclear how the electron doping of $Y_2C_3$ can be realized. At present, there are no detailed first-principles calculations of the two carbides, $Y_2C_3$ and $YC_2$, which could throw light on these questions. In any case, further theoretical and experimental studies of complex metal compounds with light elements are of great interest.

To conclude, our all-electron linear-response calculations of the phonon-related properties of some transition metal carbides and nitrides in the cubic structure confirm the theoretical results obtained by other methods. For the best-investigated compound NbC a rather good agreement with available experimental data is found. Based on our first-principles results, we demonstrate that the main reason for relatively low $T_c$ values, which is, contrary to the expectations, observed in these compounds, is a small contribution of the optical phonons (which are mostly related to the light element atoms) to the electron–phonon coupling. This is, in turn, related to a low density of electron states of the light element at the Fermi level. Thus, a promising way to reach higher $T_c$ is the search for compounds in which the chemical potential falls within an energy range where the electron states of the light element are sufficiently high. In this case these electron states would efficiently interact with high-frequency optical phonons.

In a sense, yttrium carbide provides an example, since its chemical potential falls within a peak of $N_C(0)$. Unfortunately, the carbon states at $E_F$ are virtually pure $p$-states and, according to the selection rule ($l \to l \mp 1$), the EPI matrix element of YC is small. For the Hopfield parameter (15) to be large, it is necessary that the $p$-states of light element would be hybridized, say, with $d$-states of transition element or with own $s$-states like in $MgB_2$. Such hybridization is most likely to occur in complex compounds similar to $Y_2C_3$, rather than in monocarbides and mononitrides.

Recent development of new high-pressure techniques for electric and magnetic measurements in diamond-anvil cells makes possible the investigation of superconductivity in a megabar pressure range. For example, pressure-induced changes in `phonon frequencies,` electron–phonon coupling constant and superconducting temperature of HfN, ZrN, and NbN have been recently studied[29] up to 30 GPa by means of Raman-scattering measurements. As mentioned above, stoichiometric NbN is dynamically unstable in the cubic NaCl-type structure. Our calculations of the phonon spectra and Eliashberg function for cubic HfN and ZrN at high pressure are underway and will be reported elsewhere.

## Acknowledgments

The authors are grateful to M. I. Katsnelson and S. Q. Wang for the fruitful discussions. This work was supported by the Leading Scientific Schools Program, Presidium and the Branch of General Physical Sciences, Russian Academy of Sciences, the RFBR grants nos. 05-02-17359 and 06-02-16978, the NWO–RFBR grant no. 047.016.005, and the RFBR–GFEN grant no. 05-02-39012. S.V.E. acknowledges the Landau Stipendium (Jűlich, Germany). M.V.M. is grateful to the University of Twente (Enschede, the Netherlands), where part of this work has been done.

# References

[1] *Problema Vysokotemperaturnoi Sverkhprovodimosti*, edited by V. L. Ginzburg and D. A. Kirzhnits (Nauka, Moscow, 1977) [*High-Temperature Superconductivity*, edited by V. L. Ginzburg and D. A. Kirzhnits (Consultants Bureau, 1982)].

[2] L. E. Toth, *The Transition Metal Carbides and Nitrides* (Academic Press, New York, 1971).

[3] W. E. Pickett, B. M. Klein, and D. A. Papaconstantopoulos, Physica **108,** 667 (1981); D. A. Papaconstantopoulos, W. E. Pickett, B. M. Klein, and L.L. Boyer, Nature **308,** 494 (1984).

[4] B. M. Klein and D. A. Papaconstantopoulos, Phys. Rev. Lett. **32,** 1193 (1974).

[5] Y. Zhoa and S. He, Solid State Commun. **45,** 281 (1983).

[6] G. Linker, R. Smithey, and O. Meyer, J. Phys.: Met. Phys. **14**, L115 (1984)

[7] K. Inumaru, K. Baba, and S. Yamanaka, Phys. Rev. B **73**, 052504 (2006)

[8] E. G. Maksimov and Yu. A. Shilov, Usp. Fiz. Nauk **169**, 1223 (1999) [Phys. Usp. **42**, 1121 (1999)].

[9] E. G. Maksimov and D. Yu. Savrasov, Solid State Commun. **119,** 569 (2001).

[10] E. G. Maksimov and O. A. Pankratov, Usp. Fiz. Nauk **116**, 385 (1975) [Sov. Phys. Usp. **18**, 481 (1976)].

[11] N. W. Ashcroft, Phys. Rev. Lett. **92,** 187002 (2004).

[12] J. Nagamatsu, N. Nagakawa, T. Muranaka, Y. Zenitani, and J. Akimitsu, Nature **410,** 63 (2001).

[13] S. Y. Savrasov, Phys. Rev. B **54,** 16470 (1996).

[14] S. Y. Savrasov and D. Y. Savrasov, Phys. Rev. B **46,** 12181 (1992).

[15] E. G. Maksimov, M. V. Magnitskaya, S. V. Ebert, and S. Yu. Savrasov, Pis'ma Zh. Eksp. Teor. Fiz. **80,** 623 (2004) [JETP Lett. **80,** 548 (2004)].

[16] E. I. Isaev, A. Ahuja, S. I. Simak, A. I. Lichtenstein, Yu. Kh. Vekilov, B. Johansson, and I. A. Abrikosov, Phys. Rev. B **72**, 0645515 (2005)

[17] R. Bauer, A. Y. Liu, and D. Strauch, Physica B **263–264,** 452 (1999).

[18] S. Yu. Savrasov and E. G. Maksimov, Usp. Fiz. Nauk **165**, 773 (1995) [Phys. Usp. **38**, 737 (1995)].

[19] E. G. Maksimov, D. Yu. Savrasov, and S. Yu. Savrasov, Usp. Fiz. Nauk **167**, 353 (1997) [Phys. Usp. **40**, 337 (1997)].

[20] P. Hohenberg and W. Kohn, Phys. Rev. B **136,** 864 (1964).

[21] W. Kohn and L. J. Sham, Phys. Rev. A **140,** 1133 (1965).

[22] E. G. Maksimov, Usp. Fiz. Nauk **170,** 1033 (2000) [Sov. Phys. Usp. **43**, 965 (2000)].

[23] U. von Barth and L. Hedin, J. Phys. C **5**, 1629 (1972).

[24] P. Blőchl, O. Jepsen, and O. K. Andersen, Phys. Rev. B **49**, 16 223 (1994).

[25] H. G. Smith and W. Gläser, in *Proceedings of the International Conference on Phonons*, edited by M. A. Nusimovici (Flammarion, Paris, 1971).

[26] C. Y. Allison, F. A. Modine, and R. N. French, Phys. Rev. B **35,** 2573 (1987).

[27] P. Blaha, K. Schwarz, G. K. H. Madsen, D. Kvasnicka, and J. Luitz, *WIEN2k: An Augmented Plane Wave + Local Orbitals Program for Calculating Crystal Properties*, (Karlheinz Schwarz, Techn. Universität Wien, Austria, 2001) ISBN 3-9501031-1-2.

[28] G. M. Eliashberg, Zh. Eksp. Teor. Fiz. **39**, 1437 (1961) [Sov. Phys. JETP **12**, 1000 (1961)].

[29] X.-J. Chen, V. V. Struzhkin, S. Kung, H.-K. Mao, R. J. Hemley, and A. N. Christensen, Phys. Rev. B **70**, 014501 (2004)

[30] G. L. W. Hart and B.M. Klein, Phys. Rev. B **61,** 3151 (2000).

[31] J. Chen, L. L. Boyer, H. Krakauer, and M. J. Mehl, Phys. Rev. B **37,** 3295 (1988).

[32] J. J. Hopfield, Phys. Rev. B **86,** 443 (1969).

[33] P. B. Allen, in *Dynamical Properties of Solids,* edited by G.K. Horton and A.A. Maradudin (North-Holland, Amsterdam, 1980), Vol. 3, Ch. 2.

[34] J. M. An and W. E. Pickett, Phys. Rev. Lett. **86**, 4366 (2001)

[35] K.-W. Lee and W. E. Pickett, Phys. Rev. Lett. **93**, 237003 (2004)

[36] G. Amano, S. Akutagawa, T. Muranaka, Y. Zenitani, and J. Akimitsu, J. Phys. Soc. Jpn. **73,** 530 (2004).

[37] R. J. Cava, H. Takagi, H. W. Zandbergen, J. J. Krajewski, W. F. Peck, Jr., T. Siegrist, B. Batlogg, R. B. van Dover, R. J. Felder, K. Mizuhashi, J. O. Lee, H. Eisaki, and S. Uchida, Nature **367,** 252 (1994).

[38] I. R. Shein and A. L. Ivanovskii, Solid State Commun. **139,** 223 (2004).

[39] D. J. Singh and I. I. Mazin, Phys. Rev. B **70,** 052504 (2004).

**FIGURE CAPTIONS**

FIG. 1. (a) Total electron density of states for NbC at equilibrium volume $V_0$ (solid line), $V = 0.85V_0$ (dashed line), and $V = 0.7V_0$ (dash-dotted line). (b, c) Partial *s*, *p*, and *d* DOS at Nb (b) and C (c) site at normal pressure. Energy is measured relative to the Fermi level.

FIG. 2. Eliashberg function $\alpha^2(\Omega)F(\Omega)$ for NbC at equilibrium volume $V_0$ (solid line), $V = 0.85V_0$ (dashed line), and $V = 0.7V_0$ (dash-dotted line).

FIG. 3. Calculated (solid line) and measured[26] (dashed line with circles) temperature dependence of the electrical resistivity of NbC.

FIG. 4. Calculated (solid line) and measured[26] (dashed line) reflectance of NbC.

FIG. 5. Imaginary (a) and real (b) parts of optical conductivity of NbC. Theoretical results are shown by solid line and experimental data[26] by dashed line.

FIG. 6. (a) Total electron density of states for YC. (b, c) Partial *s*, *p*, and *d* DOS at Y (b) and C (c) site. Energy is measured relative to the Fermi level.

FIG. 7. Phonon dispersions (a) and Eliashberg function (b) for YC at normal pressure.

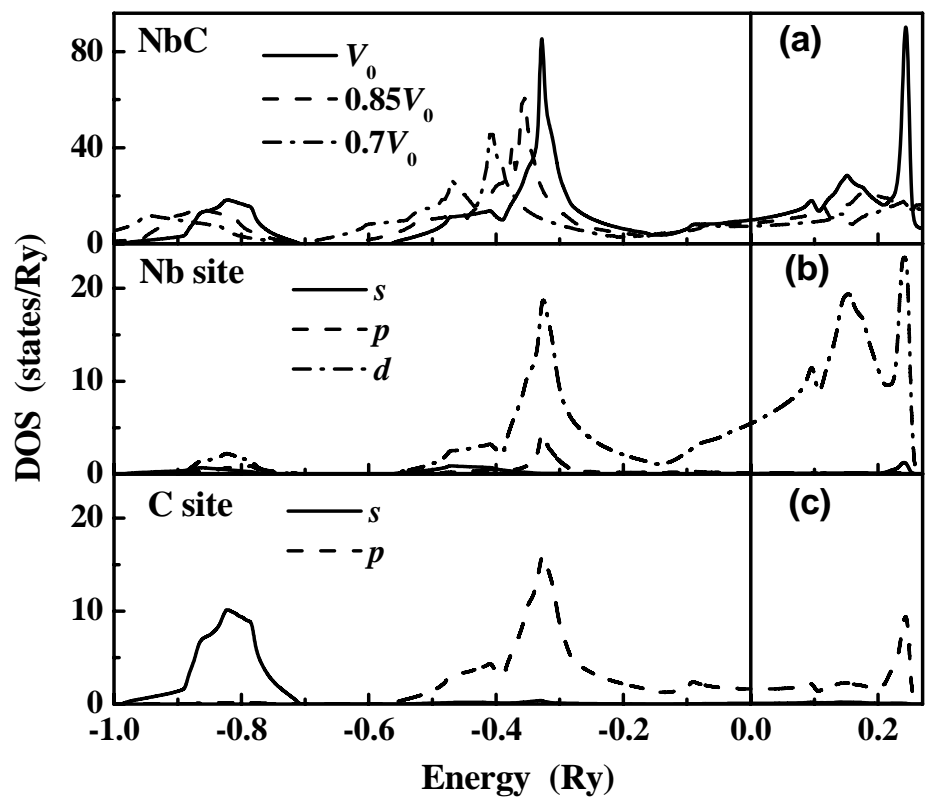

FIG. 1.

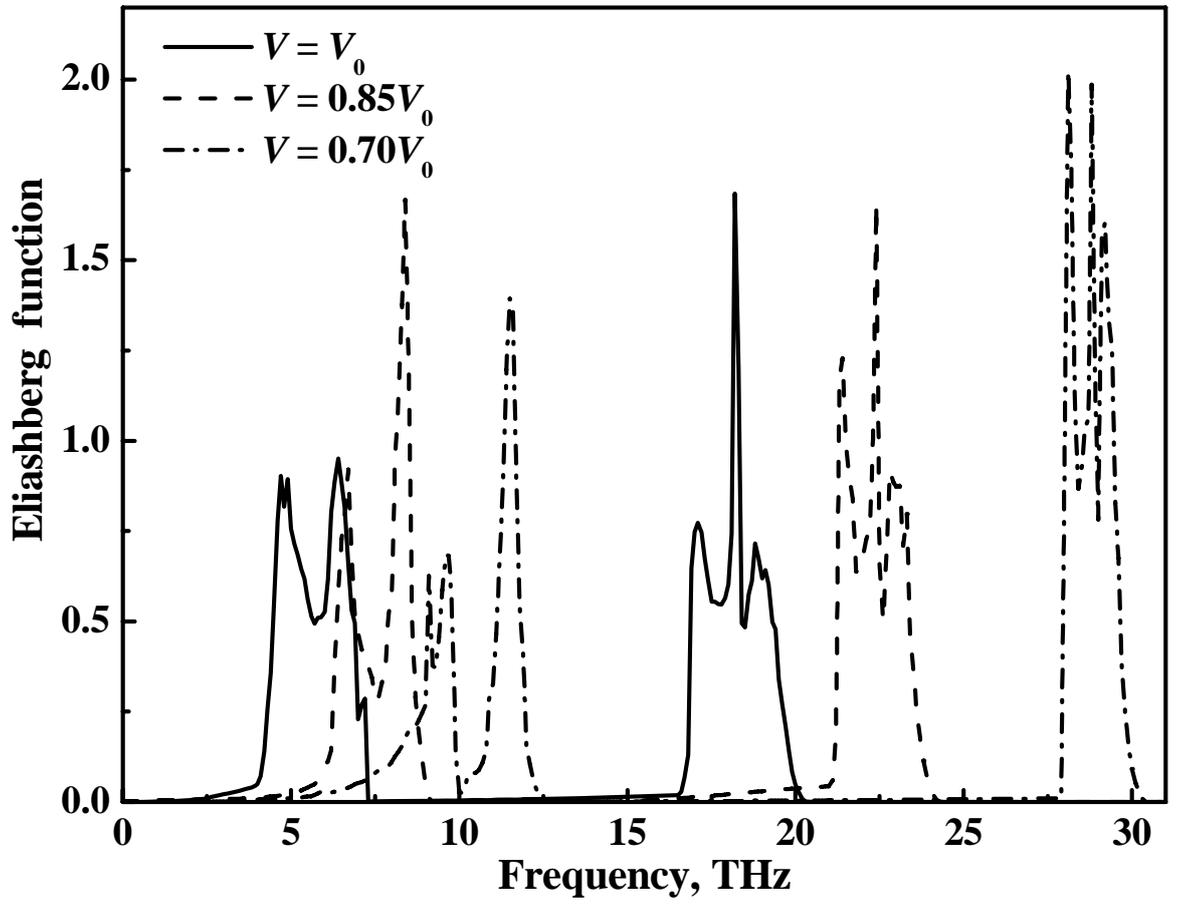

FIG. 2.

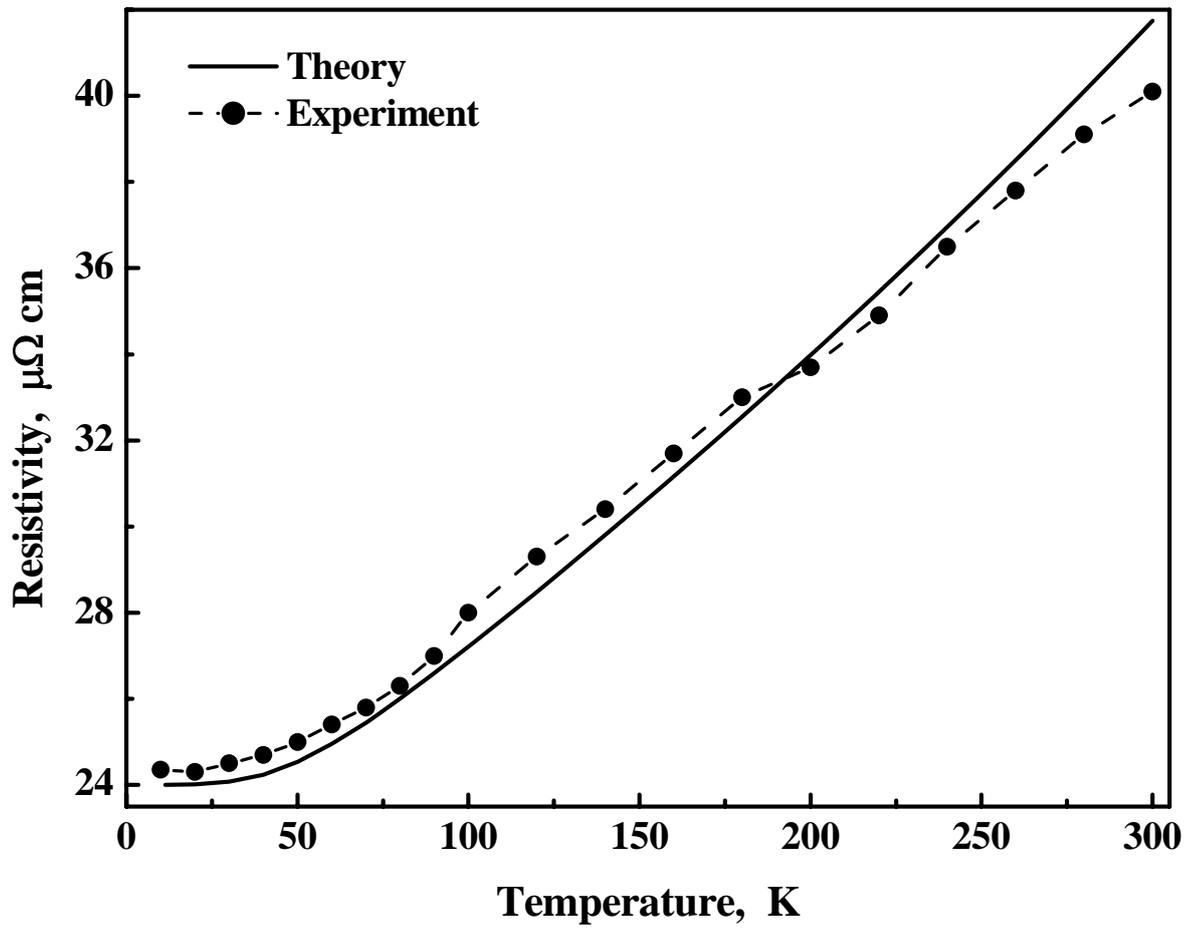

FIG. 3.

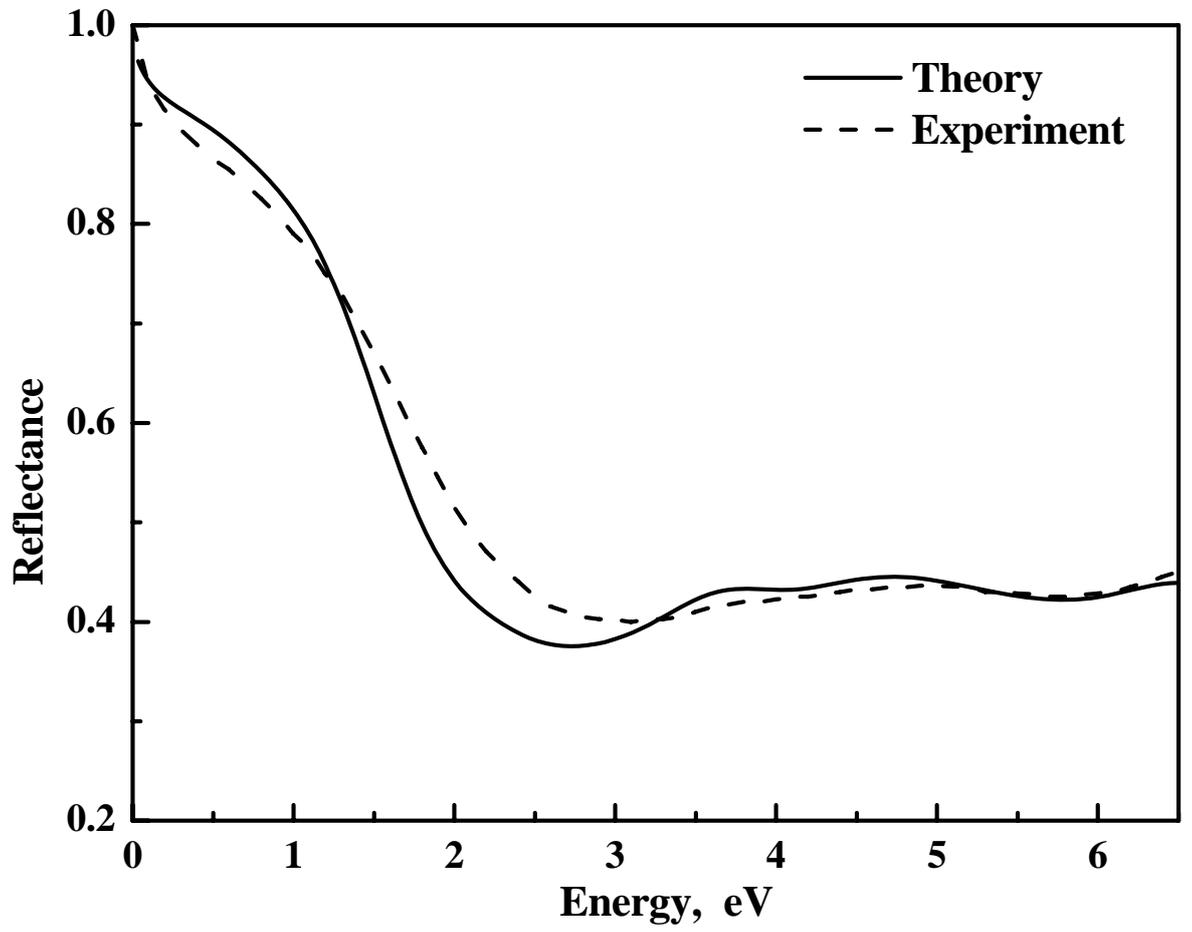

FIG. 4.

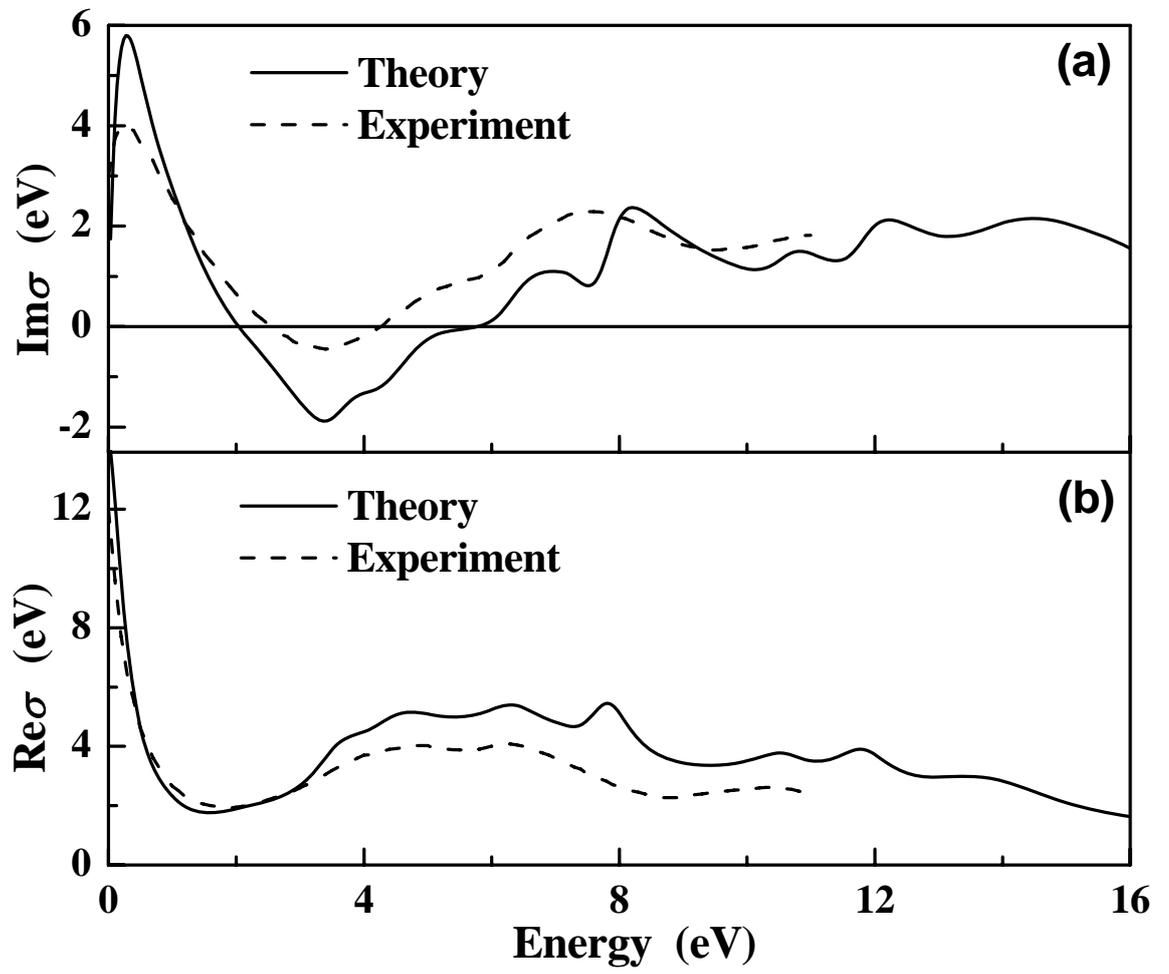

FIG. 5.

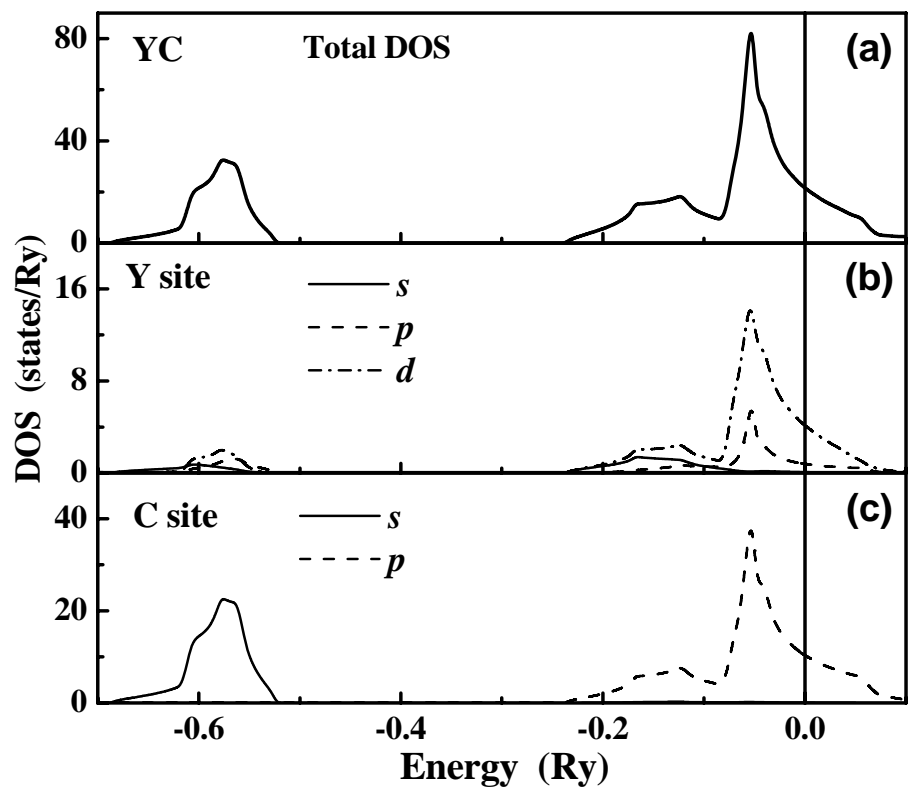

FIG. 6.

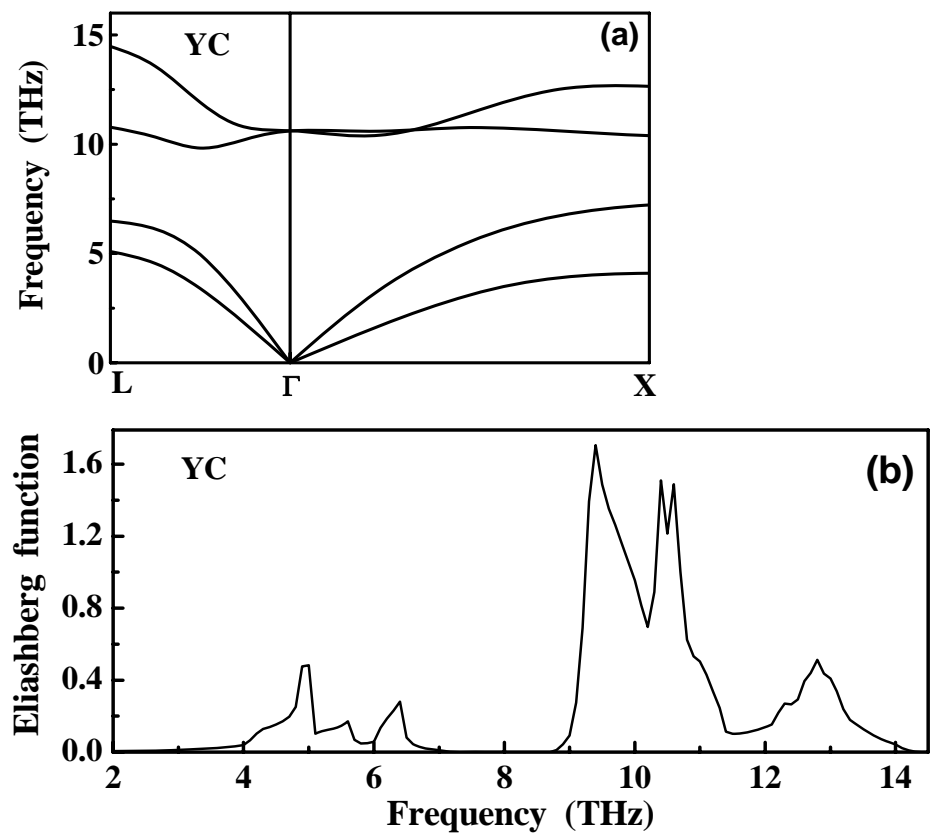

FIG. 7.